\documentclass[twocolumn,prl,showpacs,amssymb,superscriptaddress]{revtex4-1}

\usepackage{graphicx}
\usepackage{amsmath}
\usepackage{amsfonts}
\usepackage{amsthm}
\usepackage{amssymb}
\usepackage{amsbsy}
\usepackage{wasysym}
\usepackage{bm}
\usepackage{mathrsfs}
\usepackage{color}
\usepackage{hyperref}

\newcommand{\defeq}{\mathrel{\mathop:}=}

\date{\today}
\begin{document}
\title{Solvable Hydrodynamics of Quantum Integrable Systems}

\author{Vir B. Bulchandani}
\affiliation{Department of Physics, University of California, Berkeley, Berkeley CA 94720, USA}

\author{Romain Vasseur}
\affiliation{Department of Physics, University of California, Berkeley, Berkeley CA 94720, USA}
\affiliation{Materials Sciences Division, Lawrence Berkeley National Laboratory, Berkeley CA 94720, USA}
\affiliation{Department of Physics, University of Massachusetts, Amherst, MA 01003, USA}

\author{Christoph Karrasch}
\affiliation{Dahlem Center for Complex Quantum Systems and Fachbereich Physik, Freie Universit\"at Berlin, 14195 Berlin, Germany}

\author{Joel E. Moore}
\affiliation{Department of Physics, University of California, Berkeley, Berkeley CA 94720, USA}
\affiliation{Materials Sciences Division, Lawrence Berkeley National Laboratory, Berkeley CA 94720, USA}

\begin{abstract}
The conventional theory of hydrodynamics describes the evolution in time of chaotic many-particle systems from local to global equilibrium. In a quantum integrable system, local equilibrium is characterized by a local generalized Gibbs ensemble or equivalently a local distribution of pseudo-momenta. We study time evolution from local equilibria in such models by solving a certain kinetic equation, the ``Bethe-Boltzmann'' equation satisfied by the local pseudo-momentum density. Explicit comparison with density matrix renormalization group time evolution of a thermal expansion in the XXZ model shows that hydrodynamical predictions from smooth initial conditions can be remarkably accurate, even for small system sizes. Solutions are also obtained in the Lieb-Liniger model for free expansion into vacuum and collisions between clouds of particles, which model experiments on ultracold one-dimensional Bose gases.
\end{abstract}
\maketitle

\paragraph{Introduction.}

Understanding the dynamics of interacting, many-body quantum systems far from equilibrium remains one of the most challenging problems in modern physics. In recent decades, this problem has taken on a new urgency thanks to rapid progress in the experimental construction of ultracold atomic systems.  The tools available for strongly non-equilibrium dynamics with non-uniform initial conditions, even in integrable models whose equilibrium properties can be calculated exactly, have been restricted to low temperature and conformal invariance~\cite{sotiriadiscardy,bernarddoyon}, to specific quantities~\cite{VKM}, or to long-time asymptotic behaviour~\cite{Doyon,Fagotti,2016arXiv161207265D,BBH}.  Quantum integrable models include experimentally relevant examples like the Heisenberg antiferromagnet and the Lieb-Liniger gas in one dimension.  They possess extensively many conserved quantities, which prevent them from thermalizing like generic ergodic systems and often result in ballistic transport properties.

The fact that integrable models can have unusual ``generalized hydrodynamics'' due to an infinite number of local conservation laws was first understood in the context of classical particle systems~\cite{Boldrighini1983,hardrod}. This stands in contrast to conventional hydrodynamics, which describes transport of only three conserved quantities, namely mass, momentum, and energy. The generalized hydrodynamics of quantum integrable models was developed recently in studies of the non-equilibrium steady state~\cite{sotiriadiscardy,bernarddoyon,karraschilanmoore,Doyon2015190,PhysRevB.88.134301,Bhaseen:2015aa,BernardDoyonReview,VKM,PhysRevB.90.161101,PhysRevB.93.205121,1742-5468-2016-6-064005,1742-5468-2016-6-064010,2017arXiv170106620L} that is established at the junction between two infinite reservoirs~\cite{Doyon,Fagotti}.  An important insight is that making a local-density-type approximation for \textit{all} local conserved charges implies a conservation law at the level of the local pseudo-momentum distribution. Thus in the context of integrable models, the hydrodynamic equations imply a fundamental ``Bethe-Boltzmann equation'', which is an inversion of the logic familiar from conventional statistical mechanics.

The completeness of this equation for the two-reservoir steady state of the XXZ model, i.e., that the Bethe-Boltzmann equation correctly captures the physics of unusual quasilocal conservation laws~\cite{prosenxxz,PhysRevLett.111.057203,Prosen20141177,1742-5468-2014-9-P09037,PhysRevLett.115.120601,PhysRevLett.115.157201}, was tested by comparing hydrodynamic predictions to known results for spin transport in the linear-response limit~\cite{Enej,BBH}.  
It was observed at the end of~\cite{BBH} that the ansatz for two reservoirs introduced in Ref.~\cite{Fagotti} was actually valid to first order in time for arbitrary smooth, locally equilibrated initial conditions. Hydrodynamics in the two-reservoir case is a function of only one variable (say $x/t$) because of the absence of any length scale in the initial condition, and consequently a first-order solution is sufficient; every other nontrivial initial condition yields dynamics that is a function of two variables, space and time. The present work builds on this earlier observation to develop {\it converged} solutions for finite-time hydrodynamical evolution from general smooth, locally equilibrated initial conditions. 

This allows us to make novel and detailed physical predictions for finite-time dynamics in a wide range of physical systems. For example, we obtain the first practical hydrodynamic technique for the one-dimensional Bose gas~\cite{Dunjko01,Ohberg02,Pedri03,Campbell15,PhysRevA.94.051602,PhysRevA.94.063605} that applies to arbitrary local GGE initial conditions and takes into account the higher conservation laws of the underlying quantum system. This allows us to obtain detailed profiles for the evolution of the Lieb-Liniger gas from collision type initial conditions, which could, in principle, be tested in the laboratory. At the same time, our approach allows for a hydrodynamic description of finite-time spin dynamics in the XXZ chain, in excellent agreement with results obtained from density matrix renormalization group (DMRG) techniques~\cite{white92,schollwoeck,karraschdrude,barthel13,kennes16}.

\paragraph{Bethe-Boltzmann equation.} The Bethe-Boltzmann equation is a hydrodynamic description of quantum integrable systems~\cite{Doyon,Fagotti}, which aims to capture non-equilibrium dynamics in such systems using the thermodynamic Bethe ansatz (TBA)~\cite{YY,PhysRevLett.26.1301}. The Bethe-Boltzmann equation takes its simplest form for the Lieb-Liniger interacting Bose gas, which was used in~\cite{Doyon}. We briefly summarize the physical assumptions leading to this equation below, following the presentation given in~\cite{BBH} for the XXZ chain. Thus consider a one-dimensional Bose gas with delta-function interactions, placed on a line of length $L$. This has Hamiltonian
\begin{equation}
H  = \int_0^L dx \, \Psi^\dagger \left( -\frac{\hbar^2}{2m}\nabla^2  -\mu \right)\Psi + c \Psi^\dagger \Psi^\dagger \Psi  \Psi,
\label{eqLL}
\end{equation}
and the field operators satisfy canonical commutation relations $\left[ \Psi^{\dagger}(x),\Psi(y) \right]=\delta(x-y)$. It is useful to  set $\hbar=2m=1$. This system is called {\it integrable} for all values of the interaction strength $c$ because every $N$-body scattering process with $N>2$ factorizes as the product of two-body scattering processes. Integrability in this sense is reflected by the existence of infinitely many conserved charges. In a given macrostate of the Lieb-Liniger gas, with occupied density of states $\rho_k$ at pseudo-momentum $k$, these can be written as $Q_n = \int_{-\infty}^{\infty} dk\,\rho_kq_n(k)$,with $q_n(k) = k^n/n$ and $n=0,1,\ldots$. Let us now consider evolution of the Lieb-Liniger gas from {\it locally equilibrated} initial conditions; we assume that this is captured by a spatio-temporally varying pseudo-momentum distribution $\rho_k(x,t)$ \footnote{This condition is equivalent to {\it local quasi-stationarity} as defined in Ref. \cite{defect}, modulo the subtleties discussed in Ref. \cite{Pozsgay}. Thus the hydrodynamic approach is {\it a priori} inapplicable to initial states that do not admit such a description, such as N{\'e}el states.}. This yields a spatio-temporal distribution of charge density, given by
\begin{equation}
Q_n(x,t) = \int_{-\infty}^{\infty} dk\,\rho_k(x,t)q_n(k).
\label{eqcharge}
\end{equation}
We note that in order for this expression to be defined (and indeed for the hydrodynamic approach as understood in Refs. \cite{Fagotti, Doyon} to be consistent for the Lieb-Liniger gas), $\rho_k(x,t)$ must decrease more rapidly in $k$ than any power of $k$, for all $x$ and $t$~\footnote{This property is preserved under time evolution from intial conditions with compact support in $x$-$k$ space (as are required for any numerical implementation) thanks to conservation of the mode occupancies, $\int dx\,\rho_k(x,t)$.}. Motivated by conservation of $Q_n$ at the quantum mechanical level, let us postulate the local conservation law
\begin{equation}
\partial_t Q_n(x,t) + \partial_x J_n(x,t) = 0.
\label{cons}
\end{equation}
Surprisingly, the physically correct formula for $J_n(x,t)$ turns out to be given in terms of the quasiparticle velocity $v_k[\rho(x,t)]$ of collective excitations of the state with pseudo-momentum distribution $\rho_k(x,t)$, which is complicated but known from TBA~\cite{SupMat}; it has been found that~\cite{Doyon, Fagotti}
\begin{equation}
J_n(x,t) = \int_{-\infty}^{\infty} dk\,\rho_k(x,t)q_n(k)v_k[\rho(x,t)],
\label{conscurrent}
\end{equation}
in the hydrodynamic limit, which is connected to the validity of earlier conjectures for the Drude weight~\cite{PhysRevLett.82.1764}. Substituting this expression into Eq.~\eqref{cons} and appealing to completeness of conserved charges in integrable models, one deduces a conservation law for the local pseudo-momentum distribution, given by
\begin{equation}
\partial_t \rho_k(x,t) + \partial_x (\rho_k(x,t)v_k[\rho])=0.
\label{eqBB}
\end{equation}
We call this the \textit{Bethe-Boltzmann equation}, as it has the structure of a dissipationless Boltzmann equation for the occupied pseudo-momentum density. Intuitively, the Bethe-Boltzmann equation has the meaning that ``occupied quantum numbers are locally conserved''.  We emphasize that for integrable systems, the generalized hydrodynamic equations~\eqref{cons} imply the fundamental Bethe-Boltzmann equation~\eqref{eqBB}, in sharp contrast with the logic familiar from conventional statistical mechanics.

\begin{figure}[t!]
\includegraphics[width=\linewidth,clip]{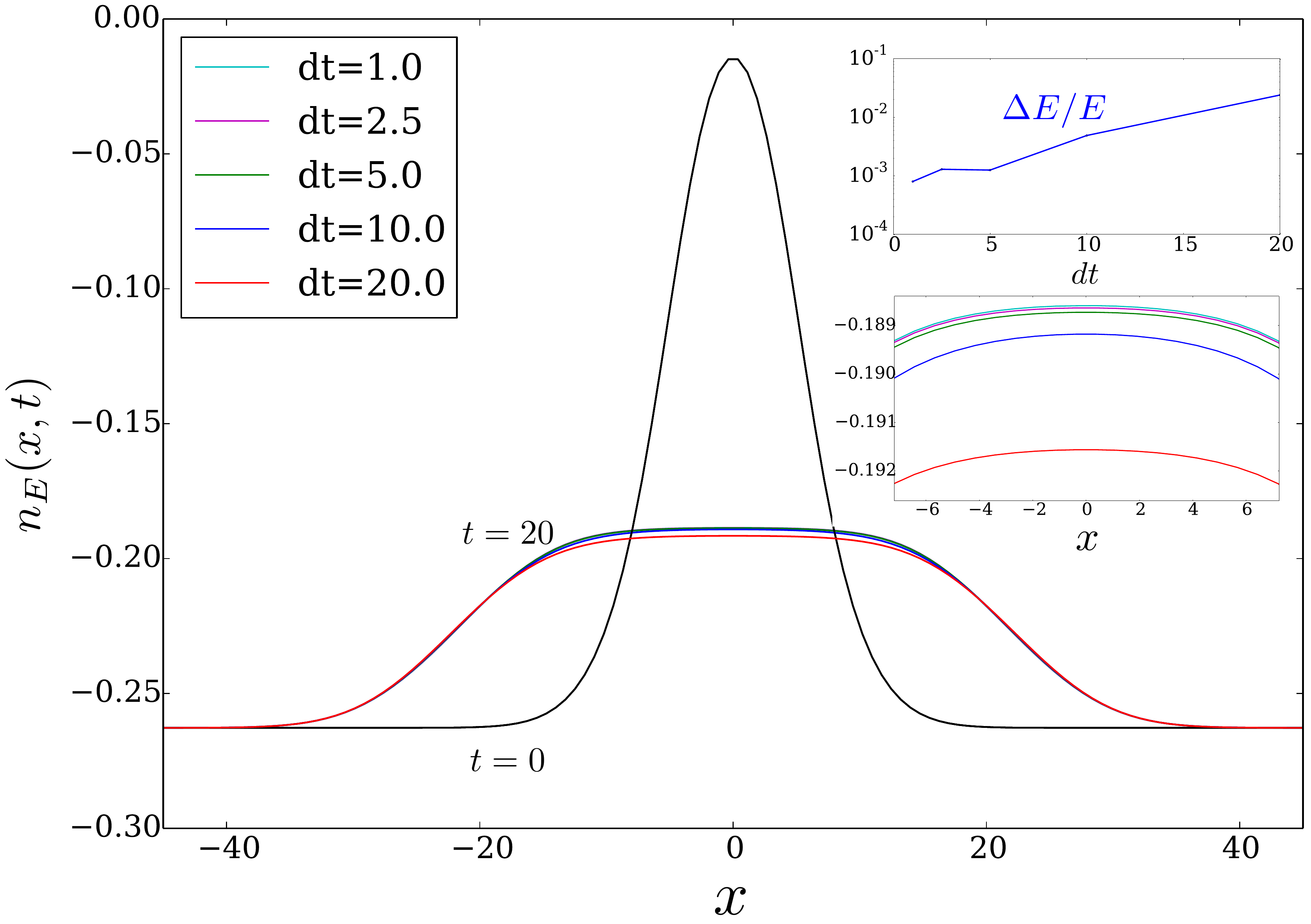}
\caption{Convergence of the method as $dt \to 0$ for an initial Gaussian temperature profile given by~\eqref{eqGaussianbeta} with $\beta_0=2.0, \beta_M=0.1$ and $L=8$ in the XXZ spin chain at $\Delta=\frac{1}{2}$. The numerical solution at time $t=20$ is rapidly converging as $dt$ is decreased, with $dt\sim10$ being already quite accurate. Insets: {\it top}: relative error in total energy, showing energy conservation as $dt \to 0$. {\it bottom}: close-up of the main figure showing convergence.}
\label{FigConvergence}
\end{figure}

 \begin{figure*}[t!]
\includegraphics[width=\linewidth,clip]{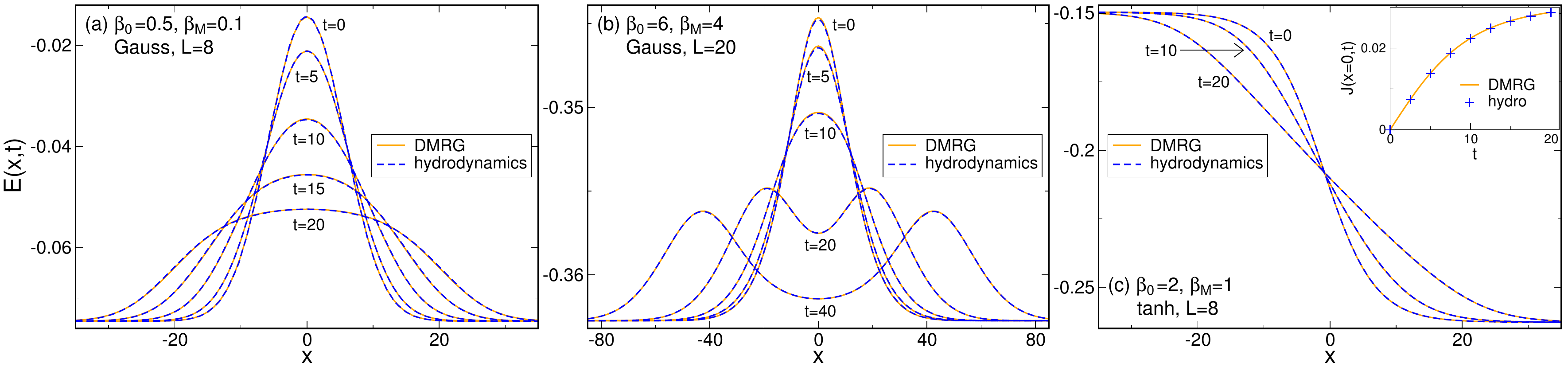}
\caption{Time evolution of energy density from various initial temperature profiles at $t=0$ for the XXZ spin chain at $\Delta=\frac{1}{2}$. {\it Left}: high temperature Gaussian initial state. {\it Middle}: low temperature Gaussian initial state. {\it Right}: two-reservoir setup with temperatures $\beta_0=2$, $\beta_M=1$ connected through a $\tanh(x/L)$ interpolation with $L=8$. {\it Inset}: Energy current at $x=0$ showing the approach to a non-equilibrium steady state at long times. }
\label{FigDMRG}
\end{figure*}

\paragraph{Finite time scheme.} 
In practice, it is useful to change variables to the local Fermi factor $\vartheta_k(x,t)$, defined as the ratio of occupied quantum numbers at pseudo-momentum $k$. This yields the \textit{advection form} of the Bethe-Boltzmann equation~\cite{Doyon,Fagotti}
\begin{equation}
\partial_t \vartheta_k(x,t) + v_k[\hat{\vartheta}]\partial_x \vartheta_k(x,t) = 0.
\label{eqBBadv}
\end{equation}
Whereas this equation has so far only been used to analyze self-similar non-equilibrium steady states whose properties depend only on $x/t$, the purpose of this letter is to illustrate how it can be solved efficiently at finite time for arbitrary initial conditions. We propose a numerical solution to Eq.~\eqref{eqBBadv}, based on a backwards implicit numerical scheme~\cite{BBH,numrep} which for time step $dt>0$, determines $\vartheta_k(x,t)$ from $\vartheta_k(x,t-dt)$ via the implicit equation
\begin{equation}
\vartheta_k(x,t) = \vartheta_k(x - v_k[\hat{\vartheta}(x,t)] dt ,t - dt). \label{eqNumScheme}
\end{equation}
This solves~\eqref{eqBBadv} up to order $\mathcal{O}(dt^2)$. We emphasize that the velocity in the right-hand side of Eq.~\eqref{eqNumScheme} depends non-linearly on the all of the Fermi factors at $(x,t)$, making Eq.~\eqref{eqNumScheme} an implicit equation that can be solved by numerical iteration. Details of our implementation are deferred to the Supplemental Material~\cite{SupMat}. Achieving convergence of numerical schemes for non-linear conservation laws in general, even in the low-dimensional setting, is known to be difficult~\cite{Leveque13}. It is therefore remarkable that the above scheme, applied to an extremely high-dimensional system \footnote{In practice, the exact dimension corresponds to the TBA discretization, typically around 4000 in our implementation.}, converges at all. Moreover, the scheme~\eqref{eqNumScheme} is found to converge quickly as $dt \to 0$, so that one can obtain accurate results even for large time steps $dt$. From the solution of Eq.~\eqref{eqBBadv}, one can readily compute physical quantities of interest (such as charge and current densities) using Eqs.~\eqref{eqcharge} and~\eqref{conscurrent}.

We note that in general, non-linear systems of equations of conservation type \eqref{eqBB} or advection type \eqref{eqBBadv} are difficult to understand analytically, because of the possibility of shock formation from smooth initial conditions. From the viewpoint of mathematical rigour, the conservation form \eqref{eqBB} is better defined, but existing analytical \cite{Lax} methods for understanding conservation laws have little practical utility in the present high-dimensional limit. Ordinarily, one can make little analytical progress with non-linear advection equations. However, somewhat surprisingly, the advection form \eqref{eqBBadv} lies in a special class of such systems which {\it are} possible to understand analytically. These are the ``semi-Hamiltonian'' or ``rich'' systems of hydrodynamic type~\cite{Dub,Tsarev,Serre,Fera}, and possess several interesting geometrical properties related to integrability (see Supplemental Material~\cite{SupMat}). 

\paragraph{Hydrodynamics for the XXZ Spin Chain.} The Bethe-Boltzmann formalism can be extended to study non-equilibrium dynamics and transport in any integrable system or integrable quantum field theory. A particularly interesting example is provided by the spin-$1/2$ XXZ chain with Hamiltonian
\begin{equation}
H = J \sum_{j} S^x_{j}S^x_{j+1}+S^y_{j}S^y_{j+1} + \Delta S^z_{j}S^z_{j+1} ,
\label{XXZham}
\end{equation}
where predictions from hydrodynamics can be compared to density matrix renormalization group (DMRG) results~\cite{white92,schollwoeck}.
Here, we set the coupling to $J=1$, and parameterize the anisotropy of the theory by $\Delta = \cos \gamma$.  The Bethe-Boltzmann formalism for the gapless phase ($-1<\Delta<1$) of this model is discussed in detail elsewhere in the literature~\cite{Fagotti,Enej,BBH} (see Supplemental Material~\cite{SupMat}). For the purposes of comparison with DMRG, we restrict to $\Delta =\frac{1}{2}$ (other values of $\Delta$ were considered in previous works~\cite{Fagotti,Enej,BBH} for non-equilibrium steady-states). We also focus on non-equilibrium energy transport, in particular the evolution of local energy density, given by $n_E(x,t) = \sum_{j=1}^{N_t} \int d\lambda\,e_j(\lambda)\rho_j(x,t,\lambda)$ in the hydrodynamic limit (see \cite{SupMat}).

To illustrate the range of validity of the method, we consider a strongly non-equilibrium example, namely the Gaussian initial temperature profile
\begin{equation}
\beta(x) = \beta_0 - (\beta_0 -\beta_M) {\rm e}^{-x^2/L^2},
\label{eqGaussianbeta}
\end{equation}
with $\beta_0 >\beta_M $. Physically speaking, this corresponds to a perturbation $\beta_M^{-1}$ of a background temperature $\beta_0^{-1}$, localized over a typical length $\sim L$. We first illustrate the convergence of our numerical scheme~\eqref{eqBBadv} by taking such a Gaussian initial state and letting the time step $dt \to 0$. This is depicted in Fig.~\ref{FigConvergence}. As $dt$ is lowered, the numerical solution at long times (say, $t=20$) converges very quickly, and remarkably, even one-step or two-step schemes (e.g. $dt=20$ or $dt=10$) yield good approximations to the converged solution.

We now compare the predictions of the Bethe-Boltzmann equation against DMRG calculations~\cite{white92,schollwoeck}, with the initial condition~\eqref{eqGaussianbeta} prepared using standard finite-temperature methods~\cite{karraschdrude,barthel13,kennes16}. This is shown in Fig.~\ref{FigDMRG}. We find an excellent agreement between DMRG and hydrodynamic results with $dt=2.5$ for quite different initial temperature profiles  --- Gaussian~\eqref{eqGaussianbeta} and $\tanh$ $\beta(x)= (\beta_0+\beta_M)/2+(\beta_0-\beta_M)/2 \times \tanh(x/L)$ functions. Provided that the initial condition is smooth enough for the DMRG and the thermodynamic Bethe ansatz calculations to agree at $t=0$, subsequent agreement at later times is essentially perfect. In fact, at low temperatures where it is hard to obtain smooth initial conditions in DMRG, the main source of error comes from slight disagreements in the initial conditions between the two approaches.

\paragraph{Hydrodynamics for the interacting Bose gas.}

\begin{figure}[t!]
\includegraphics[width = \linewidth]{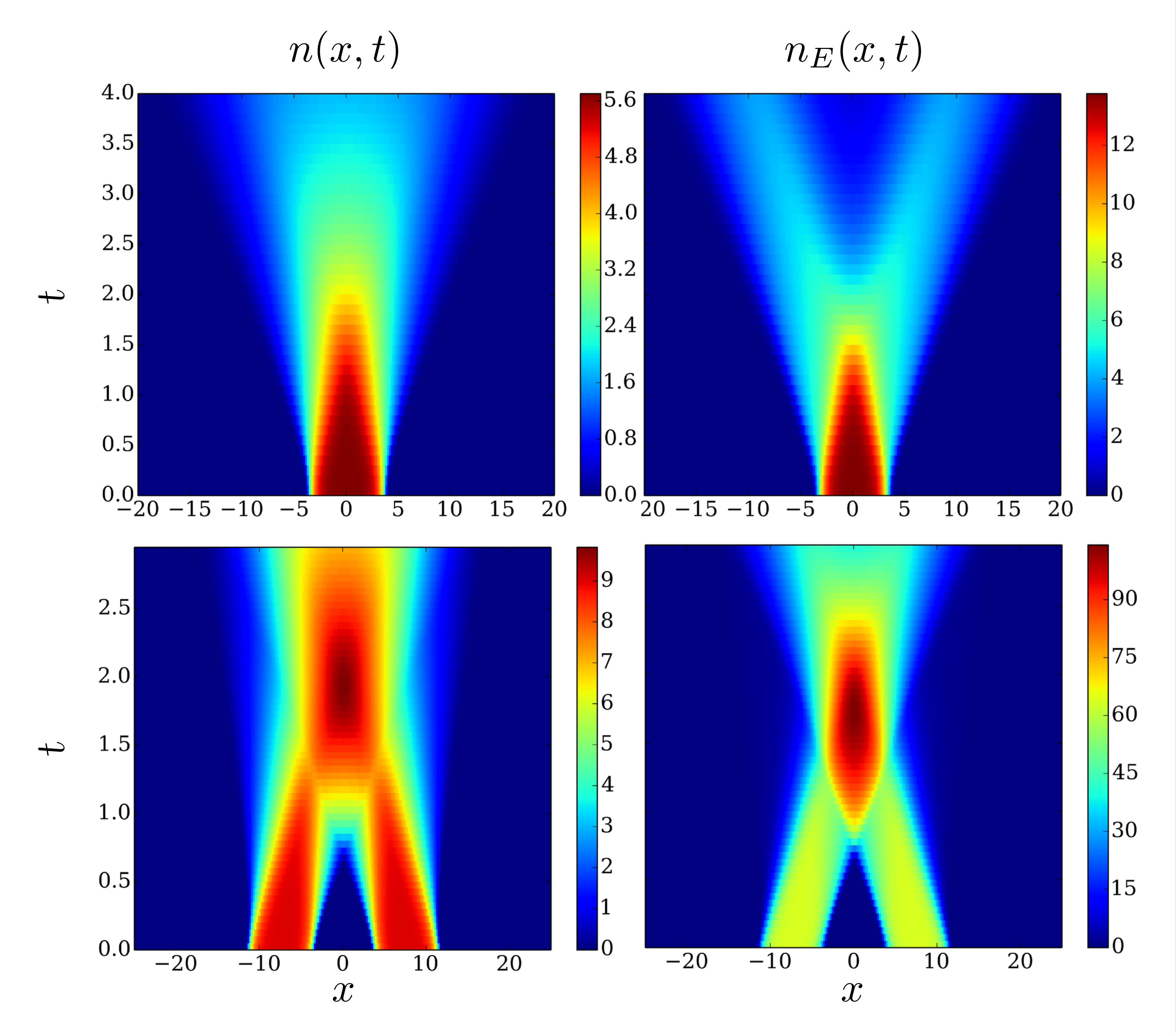}
\caption{Hydrodynamic evolutions in the Lieb-Liniger model with interaction strength $c=1$: the top panel depicts free expansion initial conditions ($dt=0.1$) and the lower panel models a collision between clouds of bosons with opposite initial momenta ($dt=0.05$).}
\label{FigExpansionBoseGas}
\end{figure}

An existing hydrodynamic description of the quasi-one-dimensional Bose gas, based on a local density approximation for the first three conserved charges of the Lieb-Liniger model, has proved effective for capturing non-equilibrium dynamics for such systems~\cite{Kolomeisky92,Kolomeisky00,Dunjko01,Ohberg02,Pedri03,Campbell15,PhysRevA.94.051602,PhysRevA.94.063605}. In its domain of physical validity~\cite{Kolomeisky00,Kheruntsyan03}, the present approach improves the existing theory by allowing for local GGE initial conditions and respecting all higher conservation laws implied by integrability. The extension of this hydrodynamic approach to other important aspects of 1D Bose gas physics, including dynamics in external potentials~\cite{tdt,shlyap,breathe,PhysRevA.94.051602,PhysRevA.94.063605,PhysRevA.95.043622} and correlation functions~\cite{Gangardt03,Kheruntsyan03,PhysRevA.95.033617,PhysRevA.95.043622}, is currently an active area of research ~\cite{GHD,GHDII}.

Two non-equilibrium quenches which are of particular experimental interest are sudden expansions of Bose gases into vacuum and collisions of clouds of ultracold bosons~\cite{PhysRevE.71.036102,Kinoshita:2006p900,PhysRevB.79.214409,PhysRevB.84.205115,bloch_expansion,PhysRevB.89.075139,PhysRevLett.109.110602,PhysRevA.85.043618,PhysRevB.88.235117,PhysRevB.95.035155,PhysRevB.95.060406}. We find that our numerical solution to Eq.~\eqref{eqBBadv} converges for initial conditions modeling both of these scenarios. From the resulting evolution in $\theta(x,t,k)$, one can track the evolution of any local conserved charge of the model. In Fig.~\ref{FigExpansionBoseGas}, we plot the time evolution of particle and energy densities (defined by $n=Q_1$ and $n_E=Q_2$ in Eq.~\eqref{eqcharge} respectively) in a Lieb-Liniger model with interaction strength $c=1$ from free expansion and collision type initial conditions. The initial states are prepared at temperature $T=1$ using a chemical potential profile interpolating between $\mu=5$ inside a box and a large negative value $\mu=-50$ outside, with the edges of the box smoothed out using tanh functions. For the collision protocol, two clouds of bosons initially prepared as in the free expansion quench are given opposite momenta $k=\pm k_0$ with $k_0=2.5$.

\paragraph{General features.} We hope that the previous examples have established that the Bethe-Boltzmann equation is a valuable tool for specific computations.  More generally, it is natural to ask which phenomena in integrable models are missed by this hydrodynamical approach and how it differs qualitatively from conventional hydrodynamics.  Clearly there is a significant assumption that the initial condition is well described by a local generalized Gibbs ensemble (GGE)~\cite{Rigol:2008kq,PhysRevLett.100.100601}. In fact, the ``thermalization problem'' for quantum integrable models, namely the question of determining the GGE to which a given quantum state converges in the long-time limit, is a difficult problem which remains unsolved in general. Nevertheless, when local equilibrium initial conditions can be imposed, hydrodynamics does seem to capture the leading behavior at long time and length scales, as illustrated in the examples above. There can also be other important subleading behaviors, beyond the approach to local equilibrium. An example is the behavior of the Lieb-Liniger model in the low-temperature limit, where it can be described by a conformal field theory or bosonization. Conformal invariance~\cite{sotiriadiscardy} and other methods~\cite{Moosavi17} both predict a Schwarzian derivative term in the time evolution from locally thermal initial conditions, which might be an example of subleading behavior beyond hydrodynamics.

An important difference between generalized hydrodynamics in integrable models and conventional hydrodynamics concerns reversibility. The collision term in the standard Boltzmann equation induces dissipation and an increase of entropy. The Bethe-Boltzmann equation is dissipationless, and in fact its time evolution is {\it reversible}. The action of microscopic time reversal on a particular state is to invert all pseudo-momenta $k$ while fixing space, so that the time-reversed pseudo-momenta are given by $\rho_k^\prime(x,t) = \rho_{-k}(x,t)$. One can show despite the complex form of the velocity $v_k$ in the Bethe-Boltzmann equations, it undergoes a simple sign change under this microscopic time reversal transformation \footnote{This follows by the dressing equations for energy and momentum, together with the fact that TBA kernels are difference kernels in the rapidity parameterization.}.

So integrable models again present some surprises compared to ordinary kinetic theory: while the description by a local GGE is certainly a great reduction in complexity compared to an arbitrary quantum state and hence irreversible, the flow in the space of local GGEs described by the Bethe-Boltzmann equation is reversible.  Presumably this means that truly diffusive behavior, as is believed to appear for example in the gapped phase of the XXZ model, lies beyond this equation; for linear-response spin transport in XXZ model, which involves both ballistic and diffusive components~\cite{sirker:2010,PhysRevB.83.035115}, hydrodynamics captures correctly the ballistic part~\cite{Enej,BBH}.

There are many possible mathematical questions regarding the existence and structure of solutions to the Bethe-Boltzmann equation~\cite{SupMat}, but we hope that the above results demonstrate its practical utility for applications to physics. It can be used as a starting point for comparison for other methods for quantum dynamics, or for incorporating integrability-breaking terms or driving.  The hydrodynamical theory of integrable models is one of many examples in recent years of how the old vine planted by Yang and Yang~\cite{YY} continues to bear fruit.


{\it Acknowledgements}: The authors acknowledge support from NSF DMR1507141
and a Simons Investigatorship (V. B. B. and J. E. M.), the Emmy-Noether program of the Deutsche Forschungsgemeinschaft under KA 3360/2-1 (C. K.) and the Quantum Materials Program supported by the Director, Office of Science, Office of Basic Energy Sciences, Materials Sciences and Engineering Division, of the US DOE under Contract No. DE-AC02-05CH11231 (R. V.).

\bibliography{fexp}

\eject
\bigskip\
\newpage

\onecolumngrid
{\center \large \bf Solvable Hydrodynamics of Quantum Integrable Systems: Supplemental Material}

\onecolumngrid
\appendix

\section{Thermodynamic Bethe Ansatz and the Bethe-Boltzmann Equations}
The Bethe-Boltzmann hydrodynamic approach is fundamentally based on the theory of quasiparticle excitations in thermodynamic Bethe ansatz~\cite{Fagotti,Doyon}. In this Appendix, we summarize the thermodynamic Bethe ansatz description of quasiparticle excitations of the Lieb-Liniger model and gapless XXZ model, providing explicit formulas for the quasiparticle velocities required in the main text.

\subsection{Lieb-Liniger model}
In the Lieb-Liniger model, the group velocity of quasiparticle excitations about thermodynamic equilibrium is given in terms of the dressed energy and momentum by $v(k) = E'(k)/P'(k)$, with dressed charges $Q$ related to bare charges $q$ via the integral equation 
\begin{equation}
Q'(k) + \int_{-\infty}^{\infty} dk'\,\mathcal{K}(k,k')\vartheta(k')Q'(k') = q'(k).
\label{dc}
\end{equation}
Here, $\vartheta$ denotes the Fermi factor of the underlying equilibrium state, and the Lieb-Liniger kernel is given in terms of the interaction strength $c$ by $$\mathcal{K}(k,k') = -\frac{c}{\pi}\frac{1}{c^2 + (k-k')^2}.$$ It is useful to introduce operators 
\begin{align}
\hat{K}[f](k) = \int_{-\infty}^{\infty} dk'\,\mathcal{K}(k,k')f(k'), \quad \hat{\vartheta}[f](k) = \vartheta(k)f(k).
\end{align}
Then from the bare values $p(k) = k$ and $e(k) = k^2/2$ we can take the formal inverse of Eq. \eqref{dc} (i.e. expand in a Neumann series) to yield the dressed values
\begin{align}
P'(k) =  (1+\hat{K}\hat{\vartheta})^{-1}[1](k), \quad E'(k) =  (1+\hat{K}\hat{\vartheta})^{-1}[k'](k),
\end{align}
giving rise to the formula
\begin{equation}
v(k) = \frac{(\hat{1}+\hat{K}\hat{\vartheta})^{-1}[k'](k)}{(\hat{1}+\hat{K}\hat{\vartheta})^{-1}[1](k)}.
\label{eqVel}
\end{equation}
If we now consider a ``local TBA'' type approximation~\cite{Doyon,Fagotti}, which amounts to the assumption that local energy density is well-defined~\cite{BBH}, then we can postulate a \textit{local Bethe equation}
\begin{equation}
\frac{\rho(x,t,k)}{\vartheta(x,t,k)} + \int_{-\infty}^{\infty} dk'\, \mathcal{K}(k,k')\rho(x,t,k') = \frac{1}{2\pi},
\label{eqlocalBA}
\end{equation}
at every point; this may be taken as a definition of the \textit{local Fermi factor} $\vartheta(x,t,k)$, which in turn yields the \textit{local quasiparticle velocity}, 
\begin{equation}
v(x,t,k) = \frac{(\hat{1}+\hat{K}\hat{\vartheta}(x,t))^{-1}[k'](k)}{(\hat{1}+\hat{K}\hat{\vartheta}(x,t))^{-1}[1](k)}.
\end{equation}
In summary, the Bethe-Boltzmann equation is shorthand for the hierarchy of equations
\begin{align}
\nonumber \partial_t \rho(x,t,k) + \partial_x (\rho(x,t,k)v(x,t,k)) &= 0 ,\\
\nonumber  \frac{2\pi\rho(x,t,k)}{1-2\pi\hat{K}[\rho(x,t,k')](k)} &= \vartheta(x,t,k), \\
\frac{(\hat{1}+\hat{K}\hat{\vartheta}(x,t))^{-1}[k'](k)}{(\hat{1}+\hat{K}\hat{\vartheta}(x,t))^{-1}[1](k)} &= v(x,t,k),
\label{loctba}
\end{align}
which together comprise a conservation law with self-consistently determined velocity. Using the local Bethe equations~\eqref{eqlocalBA}, we can instead change variables to the local Fermi factor $\vartheta(x,t,k)$, yielding the self-consistent system
\begin{align}
\nonumber \partial_t \vartheta(x,t,k) + v[\hat{\vartheta}(x,t)](k)\partial_x \vartheta(x,t,k) &= 0, \\
\frac{(\hat{1} + \hat{K}\hat{\vartheta}(x,t))^{-1}[k'](k)}{(\hat{1} + \hat{K}\hat{\vartheta}(x,t))^{-1}[1](k)} &= v[\hat{\vartheta}(x,t)](k).
\label{bbadv}
\end{align}
This form of the equation follows by various TBA identities~\cite{Fagotti,BBH}.

\subsection{Gapless XXZ Model}
Recall that the Hamiltonian for the spin-$1/2$ XXZ chain on $N$ sites, in zero external field, is given by
\begin{equation}
H = J \sum_{j=1}^{N-1} S^x_{j}S^x_{j+1}+S^y_{j}S^y_{j+1} + \Delta S^z_{j}S^z_{j+1}.
\end{equation}
We take periodic boundary conditions $S_N \equiv S_{N+1}$, set the coupling to $J=1$, and parameterize the anisotropy of the theory by $\Delta = \cos \gamma$.  We also assume in the following that the model is in its gapless phase, i.e. $-1 < \Delta < 1$. The Bethe-Boltzmann equation for this system is discussed in detail in \cite{Fagotti}. Here, we summarize the main results. The only material difference compared to the Lieb-Liniger gas is that one must now account for the ``strings'' of bound states appearing in the thermodynamic limit. Let us therefore define a \textit{string of type $j$} to be a an ordered pair, $(n_j,v_j)$, where $n_j$ is the number of spin-flips comprising the string and $v_j$ is its parity. Suppose that there are $N_t$ string types in total so that $j \in \{1,2,\ldots,N_t\}$, and $M_j$ strings of type $j$. Let $M$ denote the total number of spin-flips in the system. Then by definition, $\sum_{j=1}^{N_t} M_jn_j = M$. Upon fixing a string type $j$, we denote the rapidities of a given string $\alpha \in \{1,2,\ldots,M_j\}$ within that type by 
\begin{equation}
\lambda^j_{\alpha,a} = \lambda_{\alpha}^{j} + i[(m_j+1-2a)\gamma+ (1-v_j)\frac{\pi}{2}], \quad a = 1,2,\ldots,m_j,\end{equation}
where $m_j$ denotes the string length. It is useful to define functions 
\begin{align}
a(\lambda,n,v) &= \frac{\gamma v}{2\pi}\frac{\sin\gamma n}{\cosh \gamma \lambda - v\cos \gamma n}
\end{align}
and
\begin{align}
\nonumber a_j(\lambda) =& a(\lambda,n_j,v_j)\\
\nonumber T_{jk}(\lambda)  =& a(\lambda,|n_j-n_k|,v_jv_k)+ 2a(\lambda,|n_j-n_k|+2,v_jv_k) +\ldots \\
+&2a(\lambda,n_j+n_k-2,v_jv_k) + a(\lambda,n_j+n_k,v_jv_k).
\end{align}
The quasiparticle velocities in each string are given in terms of dressed energy and momentum by the expression $v_j(\lambda) =E'_j(\lambda)/P'_j(\lambda)$. It can be shown~\cite{Fagotti} that the dressed charge for any given quasiparticle excitation is related to the bare charge via the $N_t$ coupled integral equations
\begin{equation}
\Delta Q'_j(\lambda) + \sum_{k=1}^{N_t}\int_{-\infty}^{\infty}d\lambda' \, T_{jk}(\lambda-\lambda')\vartheta_k(\lambda') \sigma_k\Delta Q'_k(\lambda') = Q_j'(\lambda').
\end{equation}
where the $\vartheta_j$ denote the Fermi factor for strings of type $j$ and $\sigma_j = \mathrm{sgn}(v_j)$ (see \cite{Fagotti} for details). Then from the bare values $p_j(\lambda) = 2\pi a_j(\lambda)$, $e_j(k) = -Aa_j(\lambda)$ for the momentum and energy of string $j$, where $A = -2\pi J \sin \gamma/\gamma$, we obtain dressed momenta and energies
\begin{align}
P'_j(\lambda) =  2\pi(1+\hat{T}\hat{\vartheta}\hat{\sigma})^{-1}[\vec{a}]_j(\lambda), \quad E'_j(\lambda) =  (1+\hat{T}\hat{\vartheta}\hat{\sigma})^{-1}[-A\vec{a}']_j(\lambda).
\end{align}
One can deduce that
\begin{equation}
v_j(\lambda) =  \frac{1}{2\pi}\frac{(\hat{\sigma}+\hat{T}\hat{\vec{\vartheta}})^{-1}[-A\vec{a}']_j}{(\hat{\sigma}+\hat{T}\hat{\vec{\vartheta}})^{-1}[\vec{a}])_j}(\lambda),
\label{eqvXXZ}
\end{equation}
and there are now $N_t$ Bethe-Boltzmann equations, one for each string, with associated quasimomentum densities $\rho_j(x,t,\lambda)$. Finally, we note that the XXZ chain considered in the main text, with $J=1$ and $\Delta = 0.5$ possesses three strings in the thermodynamic limit, with parameters $(n_1,v_1) = (1,1)$, $(n_2,v_2) = (2,1)$ and $(n_3,v_3) = (1,-1)$. 
\section{Details of Numerical Implementation}
In this section, we describe the numerical scheme used in the main text to solve the Bethe-Boltzmann equation. For simplicity, we restrict our presentation to the Lieb-Liniger gas. Thus consider the advection form of the Bethe-Boltzmann equation 
\begin{equation}
\partial_t \vartheta(x,t,k) + v[\vartheta](k)\partial_x \vartheta(x,t,k) = 0.
\label{adv}
\end{equation}
As it stands, this is an infinite-dimensional system of coupled, non-linear equations. The first step in achieving a numerical solution is to discretize in the quasimomentum variable $k$. Let us therefore introduce a $k$-space cutoff $\Lambda$ and discretize the infinite-dimensional system \eqref{adv} at $N$ k-space points $-\Lambda < k_1<k_2<\ldots<k_N=\Lambda$. It is also useful to set $k_0=-\Lambda$. Then functions map to vectors and kernels map to matrices; for example, we have discrete analogues
\begin{align}
q_i = q(k_i), \quad \theta_i(x,t)&=\vartheta(x,t,k_i), \quad K_{ij} = \mathcal{K}(k_i,k_j)(k_j-k_{j-1})
\end{align}
of charges, Fermi factors and the Lieb-Liniger kernel respectively. Let us also introduce the matrix $A_{ij}(\theta) = K_{ij}\theta_j$. Then upon discretizing, derivatives of dressed charges $Q'$ are related to bare ones $q'$ via the matrix equation
\begin{equation}
\sum_j (\delta_{ij}+A_{ij}(\theta)) Q'_j(\theta)  = q'_j
\end{equation}
It is useful to introduce the dressing operator $U_{ij}(\theta) = (1+A(\theta))^{-1}_{ij}$ and dressed derivatives of energy and momenta $E'_i(\theta) = \sum_j U_{ij}(\theta)k_j$, $P'_i(\theta) = \sum_j U_{ij}(\theta)$. Then the discretized quasiparticle velocities can be written as $v_i(\theta) = E'_i(\theta)/P'_i(\theta)$. In terms of these functions, the discretized advection equation \eqref{adv} reads
\begin{align}
\partial_t \theta_i + v_i(\theta)\partial_x \theta_i = 0, \quad i=1,2,\ldots,N
\label{disc}
\end{align}
We note that to derive the results of the main text, we used a Simpson's rule method to compute the action of the dressing operator in evaluating the $v_i$. For large $N$, this matches the simpler discretization described here. Now consider the initial value problem for the equation \eqref{disc}, with initial condition $\theta_i(x,0) = \phi_i(x)$ at $t=0$. We propose the following backwards implicit scheme:
\begin{enumerate}
\item Choose a time step $dt > 0$.
\item At step $n=0$, set $\theta^0_i(x) = \phi_i(x)$.
\item For steps $n \geq 1$, solve for $\theta^n_i(x)$ satisfying the implicit equation $\theta^n_i(x) = \theta^{n-1}_i(x-v_i[\theta^n]dt)$ by numerical iteration.
\item Identify $\theta_i(x,n\cdot dt)$ with $\theta_i^n(x)$.
\end{enumerate}
This was shown to define a first-order numerical scheme for \eqref{disc} in a previous work \cite{BBH}. In practice, some level of spatial discretization is also necessary, and to derive the results of the main text, we applied the scheme at time step $n$ to a cubic spline interpolation of the data at step $n-1$. This ensured accuracy of the past light cone traced out by the backwards scheme at each time step.
\section{Semi-Hamiltonian Structure of Bethe-Boltzmann Hydrodynamics}
\subsection{Definition and Properties}
Rather surprisingly, the finite discretizations \eqref{disc} of the Bethe-Boltzmann equation define a ``semi-Hamiltonian'' or ``rich'' system of non-linear equations. To see this, recall that a \textit{semi-Hamiltonian system of hydrodynamic type}~\cite{Dub,Tsarev,Serre,Fera} is a system of coupled non-linear PDEs 
\begin{align}
\partial_t \theta_i + v_i(\theta_1,\theta_2,\ldots,\theta_N)\partial_x \theta_i = 0, \quad i=1,2,\ldots,N,
\label{ivp}
\end{align}
whose characteristic velocities satisfy the properties
\begin{enumerate}
\item $\partial_j v_j = 0$ for $j=1,2,\ldots,N$
\item $\partial_k\left[\partial_j v_i/(v_j-v_i)\right] = \partial_j\left[\partial_k v_i/(v_k-v_i)\right]$ for all $i\neq j\neq k \neq i$,
\end{enumerate}
where $\partial_j v_i \defeq \partial v_i/\partial \theta_j$. If the system \eqref{ivp} is semi-Hamiltonian in this sense, then it possesses infinitely many flows commuting with the velocity field $\mathbf{v}$. This leads to a range of surprising geometrical and analytic properties \cite{Tsarev,Serre}. Remarkably, it can be shown that the system \eqref{disc} is semi-Hamiltonian for any finite discretization length $N$. This is a non-trivial consequence of the structure of excitation dressing in TBA, and we briefly sketch its derivation. Let $\mathcal{S} \subset \mathbb{R}^N$ denote an open set on which the dressing operator $U$ exists. Defining the operator $\alpha_{ij}(\theta) = -\sum_{l}U_{il}(\theta)K_{lj}$, it can be shown that on $\mathcal{S}$,
\begin{equation}
\partial_j P'_i(\theta) = \alpha_{ij}(\theta)P'_j(\theta)
\label{key}
\end{equation}
and similarly for $E'$. From this, one may deduce that $\partial_j v_j = 0$, together with the formula
\begin{equation}
\partial_j v_i/(v_j -v_i) = \alpha_{ij}P'_j/P'_i =\partial_j P'_i/P'_i = \partial_j \log P'_i.
\label{deriv}
\end{equation}
By an identical argument $\partial_k v_i/(v_k -v_i) =  \partial_k \log P'_i$, and the second part of the semi-Hamiltonian property follows by commutativity of partial derivatives. In view of the theory of semi-Hamiltonian systems, we have actually shown slightly more. Recall that such systems have infinitely many flows $\mathbf{w}$ commuting with $\mathbf{v}$ and given by solutions to the system of linear equations
\begin{equation}
\partial_j w_i/(w_j-w_i) = \partial_j v_i/(v_j -v_i), \quad j\neq i,
\label{flow}
\end{equation}
with no summation. Since the result Eq. \eqref{deriv} above is independent of the bare energy, let us choose $N$ linearly independent vectors $\mathbf{a}^{(n)} \in \mathbb{R}^N$ such that $\mathbf{a}^{(0)}=\mathbf{p}'$ and $\mathbf{a}^{(1)} = \mathbf{e}'$. Then there are $N$ commuting flow fields $\mathbf{w}^{(n)}$ given by
\begin{equation}
w^{(n)}_i(\theta) = \left(\sum_{j=1}^N U_{ij}(\theta)a^{(n)}_j\right)/\left(\sum_{j=1}^N (U_{ij}(\theta)a^{(0)}_j)\right),
\end{equation}
which satisfy \eqref{flow} by our earlier reasoning. Moreover, these define linearly independent vector fields on $\mathcal{S}$ and each gives rise to a semi-Hamiltonian system in its own right. In the special case of Lieb-Liniger generalized hydrodynamics, a basis for these flows may be obtained from the first $N$ charges of the model, taking $a^{(n)}_i = {q'}^{(n)}_i=k_i^{n-1}$ for $n=0,1,\ldots,N-1$. This implies an intriguing geometrical relationship between the conserved charges of integrable models and the semi-Hamiltonian vector fields of their hydrodynamics, whose consequences in the limit $N \to \infty$ would be interesting to explore. It is also worth noting that semi-Hamiltonian structures arise naturally in the Bogoliubov or Whitham phase-averaging of {\it classical} integrable PDEs~\cite{Dub}.
\subsection{Solution by Quadrature}
In general, a semi-Hamiltonian structure implies the possibility of a solution ``by quadrature''~\cite{Fera}. Although this only works for a restricted class of initial conditions (those for which $\phi_j :\mathbb{R} \to \mathbb{R}$ has differentiable inverse), it is surprising to us that any exact solutions to the Bethe-Boltzmann equations exist at all. To motivate the construction, observe that we will have solved the discretized advection equation if we can find $\theta_j(x,t)$ and functions $f_j(\theta)$ solving the system
\begin{align}
\partial_x \theta_j = f_j(\theta_j)P'_j(\theta), \quad \partial_t \theta_j = -v_j(\theta)f_j(\theta_j)P'_j(\theta)
\label{2neq}
\end{align}
of $2N$ equations. One can show that the semi-Hamiltonian property is equivalent to integrability of this system. Let us see how this works in practice. Assuming that the $f_j(y)$ are nowhere vanishing, one can apply the inverse dressing operator to find that 
\begin{align}
\sum_{j=1}^N(\delta_{ij} + K_{ij}\theta_j) (d \theta_j/f_j(\theta_j)) = dx - k_idt.
\end{align}
Let us now consider the Cauchy problem for $t>0$, with initial state $\theta(x,0)=\phi(x)$ at $t=0$. Upon integrating the above relation along a path between $(0,0)$ and $(x,t)$, we obtain the implicit equation
\begin{equation}
\int_{\phi_i(0)}^{\theta_i(x,t)}\frac{d \theta_i}{f_i(\theta_i)} + \sum_{j=1}^N K_{ij}\int_{\phi_j(0)}^{\theta_j(x,t)}\frac{d \theta_j\,\theta_j}{f_j(\theta_j)} = x-k_it.
\label{quad}
\end{equation}
When the $f_j(y)$ are continuous, it follows by the chain rule and the implicit function theorem that this solves the system \eqref{2neq}, which implies the Bethe-Boltzmann equation. If, in addition, the initial conditions $\phi_j$ each possess a differentiable inverse, one can set $t=0$ and differentiate in $x$ to fix the $N$ functional degrees of freedom $f_i(\theta)$. 
\subsection{Geometrical Interpretation}
We now recall some geometrical properties of semi-Hamiltonian systems, due to Tsar{\"e}v \cite{Tsarev}. The main result is that for a semi-Hamiltonian system, there exists a diagonal pseudo-Riemannian metric tensor $\sum_{i=1}^N g_{ii}(\theta) d\theta_i \otimes d\theta_i$ on the space of states, satisfying
\begin{equation}
\partial_j \log \sqrt{g_{ii}(\theta)} = \partial_j v_i(\theta)/(v_j(\theta) -v_i(\theta)), \quad j\neq i.
\label{metric}
\end{equation}
To derive this, one defines first defines coefficients of a symmetric connection according to
\begin{equation}
\Gamma^i_{ij}(\theta) = \partial_j v_i(\theta)/(v_j(\theta) -v_i(\theta)), \quad j\neq i.
\end{equation}
The semi-Hamiltonian property then reads
\begin{equation}
\partial_k \Gamma^i_{ij}(\theta) = \partial_j \Gamma^i_{ik}(\theta), \quad i\neq j \neq k \neq i,
\end{equation}
which is an integrability condition for the metric compatibility condition $\partial_j \log \sqrt{g_{ii}(\theta)} = \Gamma^i_{ij}(\theta)$. This corresponds to the vanishing of certain components of the Riemann curvature tensor. For Lieb-Liniger hydrodynamics, we can write Eq. \eqref{metric} in terms of the total density of states $\rho^t$, as
\begin{equation}
\partial_j \log \sqrt{g_{ii}(\theta)} = \partial_j \log(\rho^t_i(\theta)).
\end{equation}
A solution is given by the metric tensor
\begin{equation}
g(\theta) = \sum_{i=1}^N (\rho^t_i(\theta))^2 d\theta_i \otimes d\theta_i
\end{equation}
on $\mathcal{S}$. One can check that this is flat at the free Fermion point $c =\infty$, but not generically elsewhere. Thus interactions give rise to curvature on state space, which prevents Bethe-Boltzmann hydrodynamics from being Hamiltonian in the conventional sense~\cite{Tsarev}\footnote{We were informed by B. Doyon before completion of our paper about forthcoming work on time-evolution quadrature that also points out the non-Hamiltonian property, but does not use the semi-Hamiltonian structure arising from integrability.}. In the terminology of differential geometry, the components of the total density of states define the {\it Lam{\'e} coefficients} of the state-space metric~\cite{nail}. 
\subsection{Continuum Limit}
In fact, the continuum advection equation \eqref{bbadv} appears to satisfy an infinite-dimensional analogue of the semi-Hamiltonian property with partial derivatives replaced by functional derivatives, namely:
\begin{enumerate}
\item $\frac{\delta v[\theta](k)}{\delta \theta(k)} = 0$ for all $k \in \mathbb{R}$
\item $\frac{\delta}{\delta\theta(k'')}\left[\frac{1}{v(k')-v(k)}\frac{\delta v[\theta](k)}{\delta \theta(k')}\right] = \frac{\delta}{\delta\theta(k')}\left[\frac{1}{v(k'')-v(k)}\frac{\delta v[\theta](k)}{\delta \theta(k'')}\right]$ for all $k\neq k'\neq k'' \neq k$.
\end{enumerate}
To our knowledge, the geometrical theory of such systems has not yet been developed. The first property generalizes linear degeneracy in the sense of Lax \cite{Lax}, and when combined with monotonicity, implies existence and uniqueness for solutions to a certain self-consistent ansatz proposed for two-reservoir steady states~\cite{Doyon,Fagotti}. The second property again appears to give rise to a solution by quadrature. To see this, one can simply take the continuum limit of \eqref{quad}, yielding
\begin{equation}
\int_{\phi(0,k)}^{\vartheta(x,t,k)}\frac{d \theta}{f(\theta,k)} + \int_{-\infty}^{\infty} dk'\, \mathcal{K}(k,k')\int_{\phi(0,k')}^{\vartheta(x,t,k')}\frac{d \theta\,\theta}{f(\theta,k')} = x-kt.
\end{equation}
These solutions to the Bethe-Boltzmann equation are parameterized by a functional degree of freedom $f(\theta,k)$, in two variables. Differentiating and using invertibility of the TBA kernel, one can check that this solves the system
\begin{align}
\partial_x \vartheta(x,t,k) = f(\vartheta(x,t,k),k)P'[\vartheta](k), \quad \partial_t \vartheta(x,t,k) = -v[\vartheta](k)f(\vartheta(x,t,k),k)P'[\vartheta](k),
\end{align}
which implies the continuum Bethe-Boltzmann equation. We emphasize that this exact solution is, at present, mainly of formal interest. In particular, it is logically independent from the method used to derive the results of the main text, whose details are described in a previous work~\cite{BBH}.

\end{document}